\documentclass[a4paper, amsfonts, amssymb, amsmath, reprint, showkeys, nofootinbib, twoside,pra]{revtex4-1}
\usepackage[english]{babel}
\usepackage[utf8]{inputenc}
\usepackage[colorinlistoftodos, color=green!40, prependcaption]{todonotes}
\usepackage[pdftex, pdftitle={Article}, pdfauthor={Author}]{hyperref} % For hyperlinks in the PDF
\usepackage{tabularx}
\usepackage{amsfonts}
\usepackage{graphicx,amsmath,amssymb}
\usepackage{epstopdf}
\usepackage{grffile}

\begin{document}
\date{\today }
\title{ Quantum Rabi hexagonal ring in an artificial magnetic field}
\author{Lin-Jun Li}
\affiliation{Department of Physics, and Chongqing Key Laboratory for strongly coupled Physics,  Chongqing University, Chongqing 401330, China}
\author{Li-Lu Feng}
\affiliation{Department of Physics, and Chongqing Key Laboratory for strongly coupled Physics,  Chongqing University, Chongqing 401330, China}
\author{Jia-Hao Dai}
\affiliation{Department of Physics, and Chongqing Key Laboratory for strongly coupled Physics,  Chongqing University, Chongqing 401330, China}
\author{Yu-Yu Zhang}
\email{yuyuzh@cqu.edu.cn}
\affiliation{Department of Physics, and Chongqing Key Laboratory for strongly coupled Physics,  Chongqing University, Chongqing 401330, China}
\begin{abstract}
 We present exotic quantum phases in a quantum Rabi hexagonal ring, which is derived by an analytical solution. We find that an artificial magnetic field applied in the ring induces an effect magnetic flux in the even and odd subring. It gives rise to two chiral quantum phases besides a ferro-superradiant and an antiferro-superradiant phases. With analogy to the magnetic system, two chiral phases are distinguished by the magnetization orientation in the $xy$ plane in two subrings, which correspond to skyrmion structures with different vorticity.  In such chiral phases, photons in the subrings triangle flow in the same or opposite directions by comparing to the current in the hexagonal ring, which depend on the signs of the induced magnetic flux in the subrings. Interestingly, the critical exponents of the excitation energy in two chiral phases are the same as that of the subring triangle, exhibiting subring-size dependent critical exponents. Our analysis can be straightforwardly extended to a larger lattice size with subrings of a triangular or hexagonal structure, predicting a novel universality class of superradiant phase transitions.
 An implementation of the system considered is an exciting prospect in quantum many-body simulations of light-matter interactions in future.
\end{abstract}
\maketitle

\section{Introduction}
Realizing models of exotic phases and exploring quantum phase transitions (QPT) has been an attractive objective for studying strongly interacting quantum many-body systems. A superradiant QPT of a large number
of two-level systems coupled to a bosonic field  has been attracting a remarkable amount of interest~\cite{PhysRev.93.99,Emary03,PhysRevA2008}, and now has a wide application of many recent studies in cavity~\cite{PhysRevLett2011,PhysRevLett2011dk,zhu2020} and circuit~\cite{PhysRevLett2014,PhysRevLett2014DK} quantum electrodynamics. The well-known quantum Rabi model describing the coupling between a two-level atom and a single photon exhibits interesting
integrability and novel phenomenon~\cite{braak2011,PRLIrish,chen2012,PRAzhang2016}.  Such a model exhibits a superradiant QPT in the infinite frequency limit (analogous to the thermodynamic
limit)~\cite{Ashhab2013,hwang2015,liu2017,chen2020}, which has been achieved in quantum simulations~\cite{chen2021,NCcai2021}. Such QPT in few-body systems is useful in probing a broad range of quantum phenomenon and exotic phases.

To explore intriguing quantum phenomena, much efforts have been devoted to realizing synthetic magnetic fields for bosonic excitations such as neutral atomic Bose-Einstein condensate (BEC) or cold quantum gases~\cite{bloch,bloch2012,fu2014,lin2009,dalibard2011} and photons~\cite{umucalilar2012,wang2016,roushan2017}. Synthetic magnetic fields have brought forth remarkable phenomena, such as the chiral ground-state currents of interacting photons in a 3-qubit loop~\cite{roushan2017} and fractional
quantum Hall physics in the Jaynes-Cummings Hubbard lattice~\cite{hayward2012,PhysRevA2016,Noh_2017}.  Recently, a chiral phase in the Rabi triangle has been discovered by applying a synthetic magnetic field~\cite{PhysRevLettzhang2021}. Morever, an artificial magnetic field in the light-atom coupling system can induce effective XY exchange and the Dzyaloshinskii-Moriya (DM) interactions ~\cite{PhysRevLettzhang2022}. A novel superradiant QPT induced by an artificial magnetic field shows size-dependent critical exponents in the quantum Rabi ring with $3-5$ coupled cavities ~\cite{PhysRevLettzhang2022,zhao2022anomalous}, which goes beyond the mean-field type phase transition of the conventional superradiant QPT in the Dicke model~\cite{Emary03,PhysRevA2008}.

Inspired by the size-dependent critical behaviors,
we are interested here in the emergence of exotic quantum phases and critical exponents induced by a synthetic magnetic field in the quantum Rabi hexagonal ring with six coupled cavities. We employ an analytical solution and a Bogliubove transformation to obtain the phase diagram and excitation energies accurately. We show that the next-nearest-neighbour interactions can be induced by the magnetic filed in the ring, resulting in even and odd subrings with an effect magnetic field.  There appear two chiral superradiant (CSR), a ferro-superradiant (FSR) and an antiferro-superradiant phases (AFSR) phases by using analogies to quantum magnetism. The chiral phases have a complex order paramter, and are distinguished by different magnetization orientations in $xy$ plane in the subrings triangle.  Such chiral phases are six-fold degenerate, breaking the $\mathbb{Z}_2$ and the chiral symmetries. Photons in the subrings are flowing unidirectionally in one chiral phase, while they flow in the opposite direction in the other one. It demontrates that there exist effetive magnetic fluxs with different signs in the subrings in each chiral phase. In contrast, there is no current in the FSR and AFSR phases, which only breaks the $\mathbb{Z}_2$ symmetry. The first-order QPT between the FSR (AFSR) and CSR phases is tuned by the artificial magnetic field. The scaling exponents on both sides of the chiral phase transitions are different, but are the same as that of the  subring triangle.  It demonstrate that the critical exponents become subring-size dependent as the size increases, which is different from previous ring-size dependent scaling behavior~\cite{PRLHuang2022}.

\section{Quantum Rabi hexagonal Ring }
The Hamiltonian of the Rabi ring can be given as follows
\begin{equation}  \label{Hamiltonian}
H_{RR}=\sum_{i=1}^{N}H_{R,i}+J\sum_{i=1}^{N}(e^{i\theta }a_{i}^{\dagger
}a_{i+1}+h.c.),
\end{equation}%
where $J$ is the hopping strength between neighboring resonators. The complex factor in the hopping stems from the artificial magnetic field with a flux $N\theta$~\cite{PhysRevLettzhang2021}.  The periodic boundary of the
cavity ring satisfies $a_{N+1}=a_{1}$ and $\sigma _{k}^{N+1}=\sigma_{k}^{1}$.  Each resonator interacting on-site with a two-level atom is
desribed by the quantum Rabi model as
\begin{equation}
H_{R,i}=\frac{\Delta }{2}\sigma _{z}^i+\omega a^{\dagger }_i a_i+g\left( a^{\dagger
}_i+a_i\right) \sigma _{x}^i,
\end{equation}%
where $\Delta $ is qubit energy difference, $a^{\dagger }_i$ $\left( a_i\right) $
is the photonic creation (annihilation) operator of the $i$-th single-mode cavity
with frequency $\omega $, $g\ $is coupling constants respectively, and $%
\sigma _{k}^i(k=x,y,z)$ $\ $ are the Pauli matrices. The scaled coupling
strength is $g_{1}=g/\sqrt{\Delta \omega }$. There exists a superradiant QPT in the quantum Rabi model in the infinite frequency limit $\Delta /\omega \rightarrow\infty$ ~\cite{Ashhab2013,hwang2015,liu2017,chen2020}. In this limit, each cavity undergoes such QPT.

Using the unitrary transformation $U=\Pi _{n=1}^{N}\mathtt{exp}[-ig\sigma
_{y}\left( a_{n}^{\dagger }+a_{n}\right) /\Delta ]$, we obtain the effective
Hamiltonian by projecting to the ground state of
atom $|\downarrow \rangle $
\begin{eqnarray}\label{HRR}
H_{\text{RR}}^{\downarrow } &=&\sum_{n=1}^{N}\omega a_{n}^{\dagger }a_{n}-%
\frac{g^{2}}{\Delta }\left( a_{n}^{\dagger }+a_{n}\right) ^{2}  \notag \\
&&+J\sum_{n}^{N}(e^{i\theta }a_{n}^{\dagger }a_{n+1}+h.c.)+E_{0},
\end{eqnarray}%
where the constant energy is $E_{0}=N[-\Delta /2+(\omega +3J)g^{2}/\Delta
^{2}-g^{2}/\Delta ]$, and high-order terms are dropped in the limit $\Delta
/\omega \rightarrow \infty $.

The Hamiltonian of Eq.(\ref{HRR}) can be mapped to an effective
magnetic model via a Holstein-Primakoff transformation~\cite{PhysRevLettzhang2022}. After application of $S_{n}^{z}=a_{n}^{\dagger }a_{n}-S$ and $\quad S_{n}^{+}=a_{n}^{\dagger }\sqrt{2S-a_{n}^{\dagger }a_{n}}$ with the total spin angular momentum $%
S=\Delta/2\omega $, it easily gives
approximately $S_{n}^{+}\approx \sqrt{2S}a_{n}^{\dagger }$ and $%
S_{n}^{-}\approx \sqrt{2S}a_{n}$ in the spin limit $S \rightarrow \infty$. We obtain the effective magnetic Hamiltonian
\begin{eqnarray}
H_{\mathtt{MG}} &=&\sum_{n=1}^{N}\omega S_{n}^{z}-\frac{2g^{2}}{S\Delta }%
(S_{n}^{x})^{2}+H_{XY}+H_{DM}, \\
H_{XY} &=&\frac{J_{1}\cos \theta }{S}%
\sum_{n}(S_{n}^{x}S_{n+1}^{x}+S_{n}^{y}S_{n+1}^{y}), \\
H_{DM} &=&\frac{J_{1}\sin \theta }{S}\sum_{n}\vec{e_{z}}\cdot (\vec{S}%
_{n}\times \vec{S}_{n+1}).  \label{LMGT}
\end{eqnarray}%
The XY Heisenberg interaction $H_{XY}$ which can be regarded as ferromagnetic or
antiferromagnetic depending on the sign of $J\cos \theta $. Additionally,
the Dzyaloshinskii-Moriya (DM) interaction strength in $H_{DM}$ depends on $J\sin \theta $~\cite{DZYALOSHINSKY1958241,moriya1960}, which vanishes when $%
\theta =0$ or $\pi $. The DM interactions favor noncollinear spin textures such as magnetic vortices and magnetic skymions in chiral magnates, which plays a crucial role in topological properties in magnetic systems~\cite{nagaosa2013,bogdanov1994,rossler2006,muhlbauer2009}.

It demonstrate that having a complex amplitude photon
hopping in the optical system is fundamental to obtain the DM interaction
term in the magnetic equivalent. Around $\theta =\pm \pi $ the dominant term
is the XY Heisenberg interaction, which is ferromagnetic. On the other hand, around $\theta =0$, the leading XY Heisenberg interaction is favoring anti-ferromagnetic order. The DM interaction becomes dominated  when $\theta$ approaches $\pm \pi /2$, which cause magnetization of skyrmions. There exist rich quantum phases in the effect magnetic model~\cite{PhysRevLettzhang2022}. In particular,  each magnetic phase has an equivalent phase in the quantum Rabi ring. Hence, we study the exotic quantum phases in the quantum Rabi hexagonal ring using analogies to quantum magnetism.

\begin{table}[t]\label{quantum phases}
\caption{Quamtun phases in the quantum Rabi ring for the even and odd $N$
for $\protect\theta \in \lbrack 0,\protect\pi )$. Number of different kinds of quantum  phases is listed. }%
\begin{tabularx}{0.49\textwidth} {
  | >{\centering\arraybackslash}X
  | >{\centering\arraybackslash}X
  | >{\centering\arraybackslash}X
  | >{\centering\arraybackslash}X|}
 \hline
 Even or odd $N$ & CSR ($q\neq0,\pi$)  & FSR ($q=0$) & AFSR ($q=\pi$) \\
 \hline
 $3$ & $1$  & $1$  & $0$   \\
\hline
$4$ & $1$  & $1$  & $1$   \\
\hline
$5$ & $2$ & $1$  & $0$   \\
\hline
$6$ & $2$ & $1$  & $1$   \\
\hline
$7$ & $3$  & $1$  & $0$   \\
\hline
$\cdots$ & $\cdots$ & $\cdots$  & $\cdots$   \\
\hline
$2n$ & $n-1$ & $1$  & $1$   \\
\hline
$2n+1$ & $n$ & $1$  & $0$   \\
\hline
\end{tabularx}
\end{table}

\section{Normal phase}
For a weak coupling $g_1<g_{1c}$, the ground state is a vacuum state with no excitation. It is so-called normal phase (NP). It is analogy to the paramagnetic phase where the spin is polarized by external field term along the z axis.

The quantum Rabi Hamiltonian $H_{\text{RR}}^{\downarrow }$ of Eq.(\ref{HRR}) is diagonalized exactly in the following.
By introducing  a Fourier transformation $a_k=\sum_{n=1}^N a_ne^{-ink}$, the Hamiltonian becomes $H_{\text{RR}}^{\downarrow }=\sum_{k}\omega _{k}a_{k}^{\dagger }a_{k}-\omega
g_{1}^{2}(a_{k}a_{-k}+a_{k}^{\dagger }a_{-k}^{\dagger })+E_{0}$, where the
dispersion relation is $\omega _{k}=\omega (1-2g_{1}^{2})+2J\cos (\theta -k)$
with $k=0, 2\pi/N, \cdots, 2(N-1)\pi/N$. One can give the diagonal  Hamiltonian as $H_{\text{RR}}^{\downarrow }=\sum_{k}\varepsilon _{k}a_{k}^{\dagger }a_{k}+E_{g}$, where the excitation
energy is
\begin{equation}
\varepsilon _{k}=\frac{1}{2}[\sqrt{(\omega _{k}+\omega _{-k})^{2}-16\omega
^{2}g_{1}^{4}}+\omega _{k}-\omega _{-k}].
\end{equation}%
The vanishing of the excitation energy $\varepsilon _{k}=0$ gives the critical coupling strength as
\begin{equation}\label{critical strength}
g_{1c}(k)=\frac{1}{2}\sqrt{\frac{1+4J/\omega \cos \theta \cos k+4J_{+}J_{-}}{%
1+2J/\omega \cos \theta \cos k}},
\end{equation}%
where $J_{\pm }=J/\omega \cos (\theta \pm k)$.

The critical coupling strength $g_{1c}$ depends on the momentum $k$ according to the value of $\theta$. It predicts different quantum phases emerging by tuning the flux $\theta$. For example, a CSR and a FSR phases have been found for the Rabi ring of $N=3$ and $N=4$, and an additional AFSR phase appears for the case $N=4$ ~\cite{PhysRevLettzhang2022}. Especially, in the CSR phase, the critical exponents of the excitation energy for $N=3$, $4$ and $5$ were different~\cite{PhysRevLettzhang2022,zhao2022anomalous}. It leads to a distinc universality class of the QPT with the ring-size dependent critical exponents. As $N$ increase, Table~\ref{quantum phases} displays the emerging quantum phases. Since quantum phases for $\theta\in[0,-\pi]$ is a mirror image of the one for $\theta\in[0,\pi]$.  We consider quantum phases for $\theta\geq0$. For a even $N$, there exist a FSR phase for $k=0$, an AFSR phase with $k=\pi$, and $N/2-1$ kinds of CSR phases with $k\neq 0$, respectively. While there is no AFSR phase for an odd $N$, and there exists $(N-1)/2$ kinds of CSR phases.  It is interesting to explore exotic physical properties of the emerging multiple chiral phases.
We consider $N=6$ Rabi ring with a hexagonal structure as follows.

\section{Superradiant Phases}
When $g_1$ exceeds the critical value,
the system enters the superradiant phases, for which the photon population in each cavity becomes macroscopic. Additionally, the
hopping of photons between neighboring cavities induce exotic superradiant phases according to the value of $\theta$.
For the Rabi hexagonal ring, the critical coupling strength $g_{1c}(k)$ in Eq. (\ref{critical strength}) depends on the momentum $k=0,\pm \pi /3,\pm 2\pi/3,\pi $, which predicts four superradiant phases. The phase boudary between superradiant phases are obtained by $ \theta _{c}^{\pm}=\pm\pi/2$ and
\begin{eqnarray}\label{theta}
\theta _{c}^{\pm}=\pm cos^{-1}[\mp \frac{1}{4J}\left(\omega -\sqrt{\omega
^{2}+8J^{2}}\right)].
\end{eqnarray}
 For $J/\omega =0.05$ the critical values are given explicitly as
$\theta _{c1}^{\pm }= \pm0.484\pi ,\theta _{c2}^{\pm }=\pm0.5\pi ,\theta_{c3}^{\pm }=\pm0.516\pi$.

In order to describe different superradiant phases, we incorporate
that the bosonic field acquire macroscopic occupations. We start with the
Hamiltonian in Eq.(\ref{Hamiltonian}) and displace the bosonic mode as $%
\tilde{a}_{n}= a_{n}+\alpha _{n}$ with $\alpha _{n}=A_{n}+iB_{n}$ being
complex. Making the displacements, the effective low-energy Hamiltonian is obtained by projecting to the spin subspace $\tilde{|\downarrow \rangle} $, giving
\begin{eqnarray}\label{effective hamiltonian}
H_{\text{eff}}^{\downarrow } &=&\sum_{n=1}^{N} \big[ \omega \tilde{a}_{n}^{\dagger }\tilde{a}_{n}-%
\frac{\lambda _{n}^{2}}{\Delta _{n}}\left( \tilde{a}_{n}^{\dagger }+\tilde{a}_{n}\right) ^{2}
\notag \\
&&+J\tilde{a}_{n}^{\dagger }(e^{i\theta }\tilde{a}_{n+1}+e^{-i\theta }\tilde{a}_{n-1}) \big]+E_{g},
\end{eqnarray}%
where $\Delta _{n} \equiv \sqrt{\Delta
^{2}+16g^{2}A_{n}^{2}}$, $\lambda
_{n}\equiv g\Delta /\Delta _{n}$.
The energy is obtained as $E_g=\sum_{n}\omega \alpha _{n}\alpha _{n}^{\ast }-\Delta _{n}/2+J\alpha_{n}^{\ast }(e^{i\theta }\alpha _{n+1}+e^{-i\theta }\alpha _{n-1})$.

\begin{figure}
\includegraphics[scale=0.27]{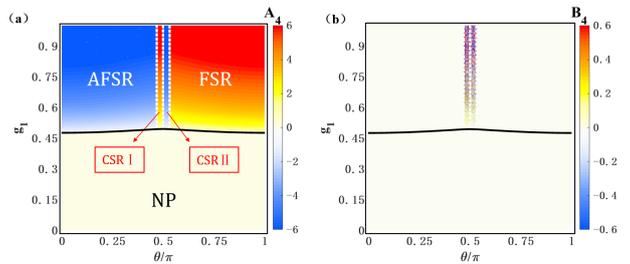}
\caption{Phase diagram in the $\protect\theta -g_{1}$ plane for the quantum
Rabi ring with $N=6$ using $A_{4}$ as an order parameter.The solid black line
represents the second-order phase boundary, while vertical dashed lines
represent the predicted first-order lines. (b) The imagirary of the order parameter $B_n$ in different quantum phases. In all our calculations, we set $%
\Delta /\protect\omega =50$ and $J/\protect\omega =0.05$, as well as choosing
$\protect\omega =1$ as the units for frequency.}
\label{phase}
\end{figure}

The order parameter $\alpha_n$ being site dependent characterizes different superradiant phases. The solutions of $\{A_{n},B_{n}\}$ are obtained by minimizing  the energy $E_g$ with respect to all $A_{n}$'s and $B_{n}$'s, which
gives%
\begin{eqnarray}
0&=&\omega B_{n}+J\sin \theta (A_{n+1}-A_{n-1})\nonumber\\
&&+J\cos \theta (B_{n+1}+B_{n-1}), \\
0 &=&\omega A_{n}-\frac{4g^{2}A_{n}}{\sqrt{16g^{2}A_{n}^{2}+\Delta ^{2}}}%
+J\cos \theta (A_{n+1}+A_{n-1})  \notag \\
&&+J\sin \theta (B_{n-1}-B_{n+1}). \label{An}
\end{eqnarray}
One easily gets $\sum_{n}B_{n}=0$ in the hexagonal ring. Moreover, we derive the equation $\omega
(B_{n}+B_{n+2}+B_{n-2})+2J\cos\theta (B_{n+1}+B_{n-1}+B_{n+3})=0$. It leads to the odd and even subrings with a triangle structure, which satisfies
\begin{eqnarray}
B_{1}+B_{3}+B_{5}=0, B_{2}+B_{4}+B_{6}=0.
\end{eqnarray}
It indicate that the nearest-neighborhood complex photon hopping can effectively induce next-nearest-neighborhood interactions with an effective magnetic field in the subrings.

Using the conditions, we give the expression as
\begin{eqnarray}  \label{Bn}
B_{n}=-J\sin \theta \frac{\omega (A_{n+1}-A_{n-1})+J\cos \theta
(A_{n-2}-A_{n+2})}{\omega ^{2}-J^{2}\cos ^{2}\theta }.  \notag \\
\end{eqnarray}
By substituting $B_{n}$ into the Eq.(\ref{An}), one obtains
\begin{eqnarray}\label{equationAn}
0 &=&A_{n}(\omega -\frac{4g^{2}}{\sqrt{16g^{2}A_{n}^{2}+\Delta ^{2}}})+J\cos
\theta (A_{n+1}+A_{n-1})  \notag \\
&&-\frac{J^{2}\sin ^{2}\theta }{\omega ^{2}-J^{2}\cos ^{2}\theta }[\omega
(2A_{n}-A_{n-2}-A_{n+2})  \notag \\
&&+J\cos \theta (2A_{n+3}-A_{n+1}-A_{n-1})].
\end{eqnarray}%
Therefore, we will look for solutions $\{A_{n},B_{n}\}$ by solving the above equation accurately.

It is interesting to understand $\{A_n,B_n\}$ using the analogies to quantum magnetism.
Using $S_{n}^{+}=(a_{n}^{%
\dagger }+\alpha _{n}^{\ast })\sqrt{2S-(a_{n}^{\dagger }+\alpha ^{\ast
})(a_{n}+\alpha )}$, one obtains $\langle S_{n}^{+}\rangle \approx \sqrt{2S-|\alpha |^2}%
\alpha ^{\ast }$. Thus the magnetic observables can be given as a magnetization vector in the xy-plane as
\begin{equation}
\vec{S_n}=(\langle \tilde S_{n}^{x}\rangle,\langle \tilde S_{n}^{y}\rangle)=(A_{n},-B_{n}),
\end{equation}
where the scaled parameters are $\langle \tilde S_{n}^{x}\rangle =\langle S_{n}^{x}\rangle/ \sqrt{2S-|\alpha |^2}$ and $\langle \tilde S_{n}^{y}\rangle =\langle S_{n}^{y}\rangle/\sqrt{2S-|\alpha |^2}$.

 Fig.~\ref{phase} shows the phase diagram of the quantum Rabi ring with the order parameter $\{A_n,B_n\}$. In the weak coupling regime $g<g_{1c}$, the order parameter $A_n$ equals to zero in the NP. As the coupling increases $g>g_{1c}$, $A_n$ grows from zero. It indicates that the system undergoes a second-order phase transition from the NP to different superradiant phases depending on $\theta$. The nonzero value of the imagary value $B_n$ shows two nontrivial superradiant phases. Four superradiant phases are characterized by $\theta_c$ as follows, which are understood better using analogies to quantum magnetism.

\textit{($1$)Ferro-superradiant phase (FSR) --}
In the FSR phase for $|\theta _{c3}^{\pm }|\leq |\theta |\leq \pi$, the order parameter $\alpha_n$ is real. $A_n$ are the same for all sites, and $B_n=0$. We obtain the analytical solution
\begin{equation}
A_n=\pm \frac{1}{4g_1}\sqrt{\frac{\Delta}{\omega}}\sqrt{\frac{16g_1^{4}}{(1+2J/\omega\cos \theta )^{2}}-1},
\end{equation}%
It demonstrate that the ground state is doubly degenerate due to the breaking of the $Z_{2}$
symmetry. It leads to the critical strength as $g_{1c}=\sqrt{1+2J/\omega\cos \theta }/2$.

\begin{figure}[tbp]
\includegraphics[trim=50 50 10 30,scale=0.28]{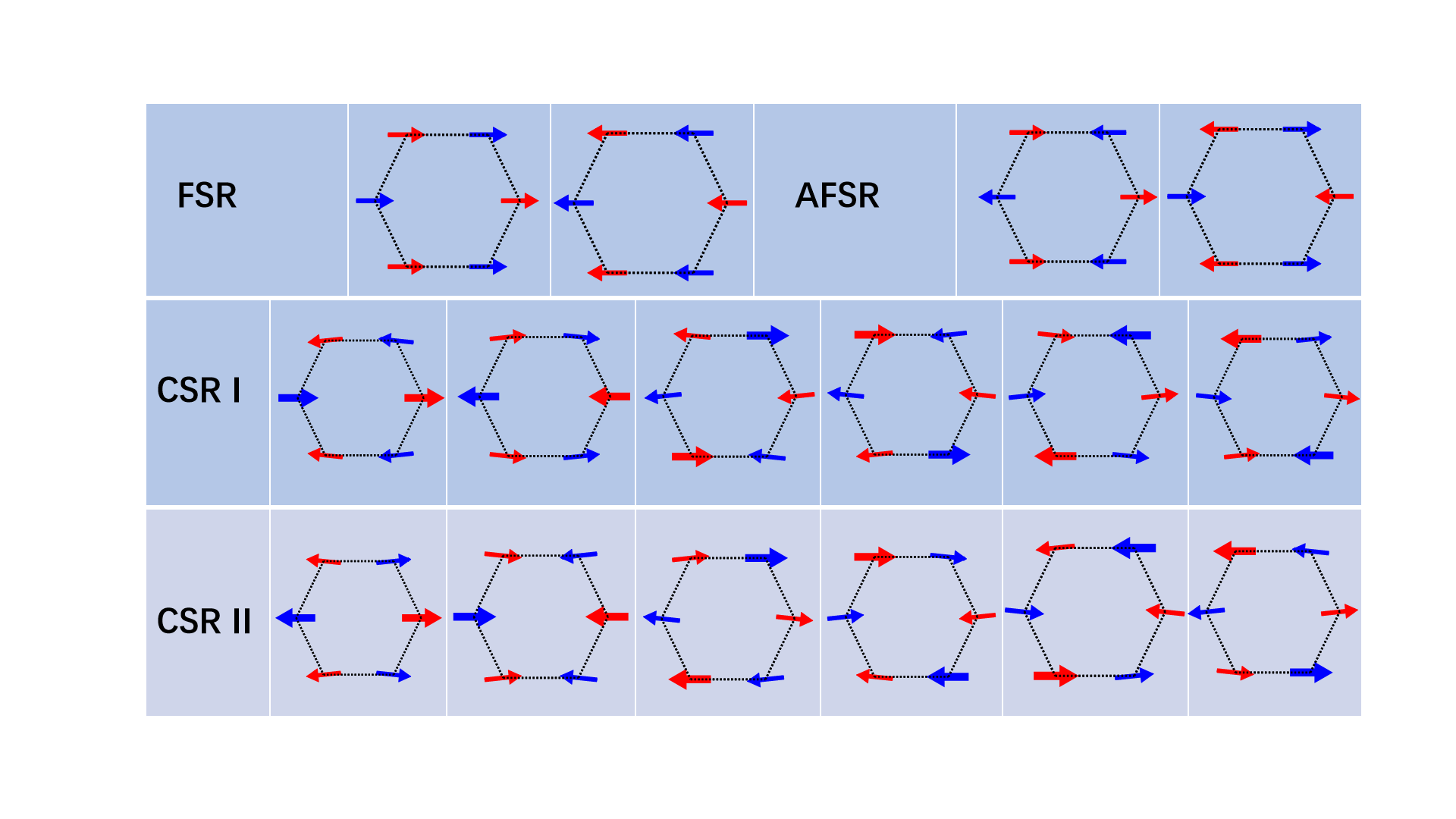}
\caption{Magnetization configurations $S_n^{x}=A_n$ and $
S_n^{y}=-B_n$ in the quantum Rabi ring in the CSR I and
CSR II phases for $N=6$ for $\protect\theta \in \lbrack 0,\protect\pi )$ in
the $xy$ plane. The arrows indicate the polarized direction of the spins for the even (blue color) and odd (red color) subrings.
The positive directions of $x$ and $y$ axises is chosen towards left and up. We choose $g_1=0.7>g_{1c}$ in the superradiant phases. }
\label{skyrmion}
\end{figure}

It is interesting to understand the degenerate configurations of the
order parameters with analogy to the spin magnetizations. In the FSR phase, the spin vector polarizes along either the $x$ or the $-x$
axis with $\langle S_{n}^{y}\rangle=0$ in Fig.~\ref
{skyrmion}, giving the spin vector $\vec{S_n}=(A_{n},0)$. It is so-called ferromagnetic superadiance.

\textit{($2$)Antiferro-superradiant phase (AFSR) --}
The AFSR phase emerges for $0\leq |\theta |\leq |\theta _{c1}^{\pm }|$. The corresponding solution is $A_{n}=-A_{n+1}=a$ and $B_{n}=0$, for which
\begin{equation}
a=\pm \frac{1}{4g_1}\sqrt{\frac{\Delta}{\omega}}\sqrt{\frac{16g_1^{4}}{(
1-2J/\omega\cos \theta )^{2}}-1 ^{2}},
\end{equation}%
It implies that the ground state is doubly degenerate. The critical coupling strength is obtained as $g_{1c}=\sqrt{%
1-2J/\omega \cos \theta }/2$.

With analogy to magnetic system, the neighboring spins are antialigned along the $x$ or the $-x$ axis in Fig.~\ref
{skyrmion}, giving the spin vector $\vec{S_n}=((-1)^na,0)$. It is so-called anti-ferromagnetic superadiance.

\textit{($3$)Chiral superradiant phase (CSR) I --}
In the chiral phase with $|\theta _{c1}^{\pm }|\leq |\theta |\leq \frac{\pi }{2}$, the order parameter $\alpha _{n}$ is complex and depends on $n$. By solving Eq.(\ref{equationAn}), one of solutions is given by
\begin{eqnarray}\label{A1}
 A_{1}=A_{4},A_{2}=A_{3}=A_{5}=A_{6},
\end{eqnarray}
and
\begin{eqnarray}\label{B1}
B_{1}=B_{4}=0,B_{2}=-B_{3}=B_{5}=-B_{6},
\end{eqnarray}
where $B_{2}=-J\sin \theta(A_2-A_{1})/(\omega -J\cos \theta)$.
There exist two degenerate solutions ( $A_{1}>0, A_2<0$) and ( $A_{1}<0,A_2>0$), which corresponds to $Z_2$ symmetry breaking. Due to the $C_3$ symmetry in the subrings, the ground state is  six-fold degenerate.

Fig.~\ref{skyrmion} shows the corresponding magnetization configurations of
spins in the CSR I phase, which is described by the spin vector $\vec{S_n}=(A_n,-B_n)$.  One observes in-plane magnetizations of skyrmions with different helicities.  Since the DM interaction in the magnetic  Hamiltonian in Eq.(\ref{LMGT}) is stronger than the XY coupling, which favors noncollinear spin structures. It yield in-plane magnetization orientation in the $xy$ plane, which is so-called chiral superradiant phase. It is observed that an pair of spins $\vec{S}_n$ in the even subring in blue color has the same orientation as $\vec{S}_{n+3}$ in the odd subring in red color in the CSR I phase. To character the skymion structure, the velocity of the skyrmion $Q$ can be expressed as $Q=\frac{1}{4\pi }\int \vec{S_n}\cdot (\partial _{x}\vec{S_n}\times
\partial_{y}\vec{S_n})dxdy$, which counts how many times $\vec{S_n}$ wraps the unit sphere~\cite{NCskymions,NTskymions2013,PhysRevBskymions}.  One can classify the skymion structures as $Q=2$ in the CSR I phase.

\textit{($4$)Chiral superradiant phase (CSR) II --}

For $\frac{\pi }{2}\leq |\theta |\leq |\theta _{c3}^{\pm }|$, the system
enters the CSR II phases. Different with the CSR I, one solution
of the solutions is
\begin{eqnarray}\label{A2}
 A_{1}=-A_{4}, A_{2}=-A_{3}=-A_{5}=A_{6},
\end{eqnarray}
and
\begin{eqnarray}\label{B2}
B_{1}=B_{4}=0, B_{2}=B_{3}=-B_{5}=-B_{6},
\end{eqnarray}
where $B_{2}=J\sin \theta (A_2+A_{1})/(\omega +J\cos \theta)$.
There exist two degenerate solutions ( $A_{1}<0, A_2<0$) and ( $A_{1}>0,A_2>0$). The chiral phase is also six-fold degenerate due to the break of both the $Z_2$ and the chiral symmetries~\cite{PhysRevLettzhang2021}.

Fig.~\ref{skyrmion} shows the magnetization configurations with the spin vector $\vec{S_n}=(A_n,-B_n)$ in the CSR II phase.
An pair of spins, $\vec{S}_n$ and $\vec{S}_{n+3}$, in the even and odd subrings has the opposite magnetization orientation, which is distinguished from that in the CSR I phase. The structures of the skymions has different vorticity $Q=1$ by comparing to the CSR I phase.

\section{Ground-state photon current}
To gain further insight into two chiral phases, we
consider the ground-state current of photons in the closed loop of the Rabi ring. Analogous to the
continuity equation in classical systems, the photon current operator is
explicitly defined as
\begin{equation}
I=i\Sigma _{n=1}^{N}\left[ (a_{n}^{\dagger
}a_{n+1}-h.c.\right] .
\end{equation}
Similarily, the photon current in the even and odd subring is given as
 $I_{\mathrm{135}}=i\left[ (a_{1}^{\dagger
}a_{3}+a_{3}^{\dagger }a_{5}+a_{5}^{\dagger }a_{1})-h.c.\right] $ and $I_{%
\mathrm{246}}=i\left[ (a_{2}^{\dagger }a_{4}+a_{4}^{\dagger
}a_{6}+a_{6}^{\dagger }a_{2})-h.c.\right] $.

\begin{figure}[tbp]
\includegraphics[trim=10 80 80 -20,scale=0.45]{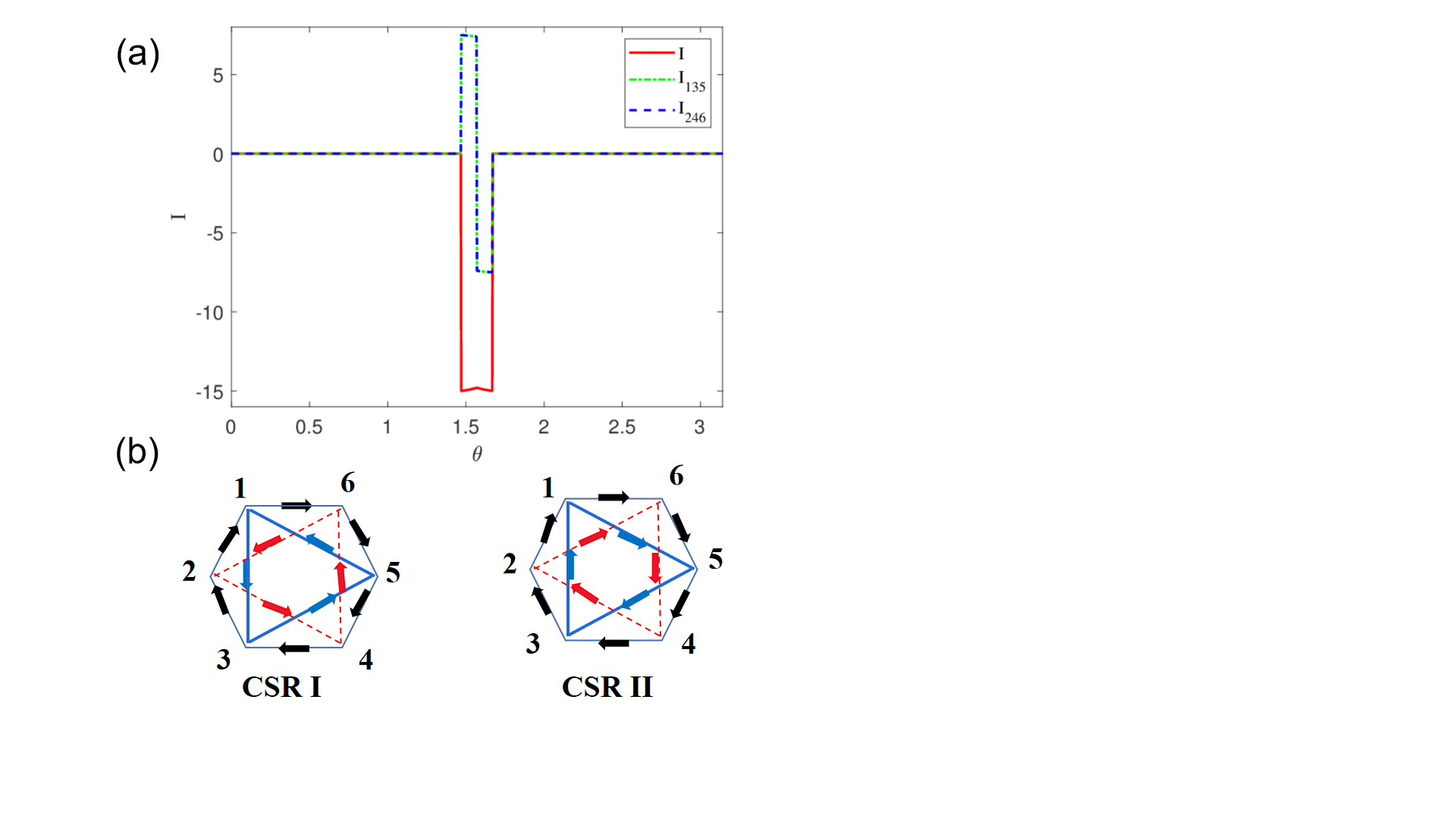}
\caption{(a)Photon current $I$ in the ring, $I_{135}$ and $I_{246}$ of the
even and odd subring in the SR regions as a function of $\protect\theta$
for $g_1>g_{1c}$. (b)Arrows represents the direction of photon current in ring (black color) and in two subrings (blue/red color) the CSR I and CSR II phases.}
\label{current}
\end{figure}

\begin{figure}[tbp]
\includegraphics[scale=0.55]{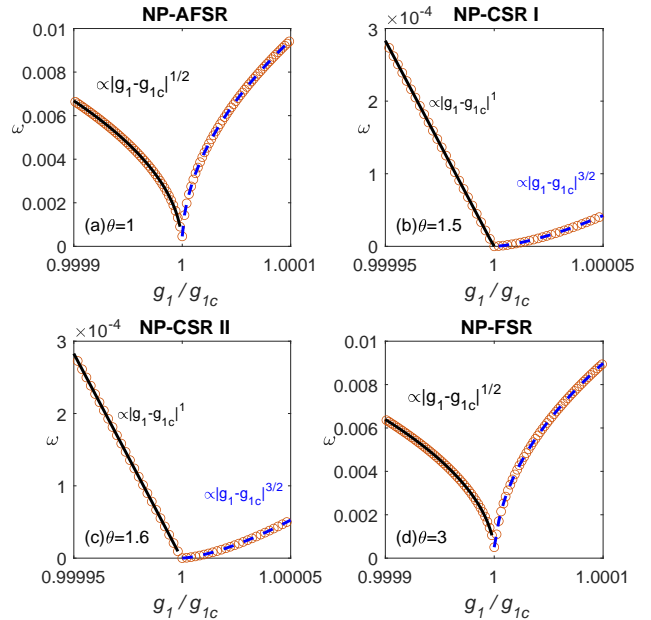}
\caption{Lowest excitation energy as a function of $g1/g1c$ across
the critical value for the NP-AFSR (a), NP-CSR I (b), NP-CSR II (c), and NP-FSR (d) phase transitions. Open markers signal the numerical results while the fitting functions are denoted by solid lines.}
\label{scaling}
\end{figure}
Using the solutions of  $\{A_n,B_n\}$ of Eqs.(\ref{A1})-(\ref{B1}) in the CSR I phase, we obtain the analytical expression of the photons current as
\begin{equation}
I=8(A_2-A_{1})B_{2},I_{\mathrm{135}}=I_{\mathrm{246}}=-I/2.
\end{equation}
The current value in the subring is half over the value in the total ring. Moreover, the direction of current in the subring is opposite to the total current.
Similarily, the current in the CSR II phase can be expressed according to Eqs.(\ref{A2})-(\ref{B2})
\begin{equation}
I=-8(A_2+A_{1})B_{2},I_{\mathrm{135}}=I_{\mathrm{246}}=I/2.
\end{equation}
 Different from the CSR I phase, we can clearly see that photons  in the even and odd subring moves towards the same direction as well as that in the Rabi ring. It demonstrates that the effective magnetic flux in the subrings has the opposite sign in two chiral phases, and is half over the total flux $6\theta$.
 We note the positive or negative next-nearest-neighbour interactions have been found in the $N=5$ Rabi ring~\cite{zhao2022anomalous}, which can be understood as a consequence of the effective magnetic field with signs in the subrings.

Fig.~\ref{current} shows the photon current in the Rabi ring as well as the even and odd subring. There is no current in the AFSR and FSR phase. However, $I$ increase sharply in two chiral phases, and behaves discontinously at $\theta_{c2}=\pi/2$, indicating a first-order transition from the CSR I to CSR II phases. Photons in the hexagonal ring moves towards clockwise direction in both chiral phases. However, in the CSR I phase the photons in the subrings flow towards anticlockwise direction, which is opposite to that in the CSR II phase. Thus, the  chiral current of photons in the subrings is adjusted by the effective magnetic field with flux $\pm 3\theta$, which is associated with the next-nearest-neighbour interactions induced by the artificial magnetic field in the ring.

\section{Excitation energy scaling}
We consider the excitation energy around the critical value to character the scaling exponents of the second-order phase transitons.
Since the Hamiltonian in Eq. (\ref{effective hamiltonian}) is bilinear in the creation and annihilation operators $a_{n}^{\dagger }$ and $a_{n}$.  It can be diagonalized as $H_{\text{eff}}^{\downarrow }=\sum_n^N \varepsilon_n b_{n}^{\dagger }b_n$ by the bosonic Bogoliubov transformation~\cite{PhysRevLettzhang2021}, where $\varepsilon_n $ is the excitation energy, the bosonic operators $
\{b_{n},b_{n}^{\dagger }\}$ are a linear combination of $\{a_{1},a_{2},a_{3},a_{4},a_{5},a_{6},a_{1}^{\dagger },a_{2}^{\dagger},a_{3}^{\dagger },a_{4}^{\dagger },a_{5}^{\dagger },a_{6}^{\dagger }\}$.

The Lowest excitation energy $\varepsilon_1 $ vanishes at the critical coupling strength of the second-order phase transition, and behaves as $\varepsilon_1\propto|g_1-g_{1c}|^{\gamma}$ across the phase boudary.
Fig.~\ref{scaling} shows the a power-law behavior of $\varepsilon_1$ when $g_1$ approaches to the critical value from above and below sides. The scaling exponent $\gamma$  is $1/2$ for the NP-AFSR and NP-FSR phase transitions in Fig.~\ref{scaling}(a) and (d). Both phase transitions belong to the same universality of the conventional superradiant phase transition in the Dicke and quantum Rabi model~\cite{hwang2015,Emary03}.

By contrast, the scaling exponents before and after the critical point are different for the NP-CSR I and NP-CSR II transitions in Fig.~\ref{scaling}(b) and (c). $\gamma$ equals to $1$ and $3/2$ for $g_1$ approaching $g_{1c}$ from below and above, respectively. The unusual critical exponents are as a consequence of the frustrated configurations in the triangle subring, which is the same as that in Rabi triangle~\cite{PhysRevLettzhang2021}.
Therefore, scaling exponents in $N=6$ Rabi ring confirms the phase transitions to be in the universality class of its subring triangle structure. Our study can be straightforwardly extended to a larger ring
size, which is associated with the subring-size dependnent critical exponents.

\section{Conclusion}
We present the analytical solution to the Rabi ring with a hexagonal structure and the phase diagram. We find an effective magnetic flux in the even and odd subrings, which is associated with the next-nearest-neighbour interactions induced by the artificial magnetic field in the ring. As the coupling strength increases, two chiral superradiant phases of the CSR I and CSR II emerge as well as a ferro-superradiant and an antiferro-superradiant phase.  In such chiral phases, one observes in-plane magnetization of skyrmions, exhibiting different velocity of the skyrmion in two chiral phases. In the CSR I phase, the current of photons in the subring has the opposite chirality by compared to the CSR II phase, which corresponds to the effective magnetic flux with different signs in the subrings. However, the scaling exponents of the excitation energy in two chiral phases are the same as that in the subring triangle case, exhibiting different critical exponents before and after the critical value. The system exhibits the subring-size dependent critical exponent, which predicts a nontrivial universality class of the Rabi ring with a subring of triangle or hexagonal structure. Studying exotic quantum phases in this few-body system would open intriguing avenues for exploring their connection to more complicated configurations.

\bibliography{refs}{}
\end{document}